\documentclass{moriond}

\def\be{\begin{equation}}
\def\ee{\end{equation}}
\def\bea{\begin{eqnarray}}
\def\eea{\end{eqnarray}}

\usepackage{ amssymb }

\newcommand{\Photo}{}

\begin{document}
\vspace*{4cm}
\title{Hint at an axion-like particle from GRB 221009A}

\author{ Giorgio Galanti }
\address{INAF, Istituto di Astrofisica Spaziale e Fisica Cosmica di Milano, Via Alfonso Corti 12, I -- 20133 Milano, Italy}

\author{ Lara Nava }
\address{INAF, Osservatorio Astronomico di Brera, Via Emilio Bianchi 46, I -- 23807 Merate, Italy}
\address{INFN, Sezione di Trieste, Via Alfonso Valerio 2, I -- 34127 Trieste, Italy}

\author{ Marco Roncadelli }
\address{INFN, Sezione di Pavia, Via Agostino Bassi 6, I -- 27100 Pavia, Italy}
\address{INAF, Osservatorio Astronomico di Brera, Via Emilio Bianchi 46, I -- 23807 Merate, Italy}

\author{ Fabrizio Tavecchio }
\address{INAF, Osservatorio Astronomico di Brera, Via Emilio Bianchi 46, I -- 23807 Merate, Italy}

\author{ Giacomo Bonnoli }
\address{INAF, Osservatorio Astronomico di Brera, Via Emilio Bianchi 46, I -- 23807 Merate, Italy}


\maketitle\abstracts{
The detection by the LHAASO Collaboration of the gamma-ray burst GRB 221009A at redshift $z = 0.151$ with energies up to $(13-18) \, \rm TeV$ challenges conventional physics. Photons emitted with energies above $10 \, \rm TeV$ at this redshift can hardly be observed on Earth due to their interaction with the extragalactic background light (EBL). We show that indeed the LHAASO Collaboration should not have observed photons with energies above $10 \, \rm TeV$ if the state-of-the-art EBL model by Saldana-Lopez et al. is taken into account. A problem therefore arises: the Universe should be more transparent than currently believed. We also show that the issue is solved if we introduce the interaction of photons with axion-like particles (ALPs). ALPs are predicted by String Theory, are among the best candidates for dark matter and can produce spectral and polarization effects on astrophysical sources in the presence of external magnetic fields. In particular, for GRB 221009A, photon-ALP oscillations occur within the crossed magnetized media, i.e. the host galaxy, the extragalactic space, the Milky Way, partially reducing the EBL absorption to a level that explains the LHAASO detection of GRB 221009A and its observed spectrum without the need of contrived choices of parameter values, which are instead compulsory within proposed emission models within conventional physics. This fact regarding GRB 221009A represents a strong hint at the ALP existence, which adds to two other indications coming from blazars, a class of active galactic nuclei.}

\section{Introduction}

The Gamma Ray Burst GRB 221009A has been detected with energies ${\mathcal E}$ up to $(13-18) \, \rm TeV$ by the LHAASO Collaboration\cite{LHAASO} at redshift $z = 0.151$\cite{redshift}, while its possible observation at $251 \, {\rm TeV}$ by the Carpet-2 Collaboration\cite{carpet} is not confirmed. The detection of GRB 221009A challenges standard physics, since at $z = 0.151$ very-high-energy photons with ${\mathcal E} > 10 \, \rm TeV$ are strongly absorbed due to their interaction with the photons of the extragalactic background light (EBL)\cite{dwek}, which are produced by stars and possibly reprocessed by the dust throughout the Universe history. In particular, current EBL models (for a list see\cite{ALPinGRB}) predict an attenuation factor of at least ${\cal O} (10^6 - 10^8)$ at ${\mathcal E} = 18 \, {\rm TeV}$.

We employ the EBL model by Saldana-Lopez et al.\cite{saldanalopez} (referred to as SL), which currently appears as the most robust, since it is based on the most recent and complete galaxy data sets and is able to minimize foreground effects (like the zodiacal light) because it is inferred by a satellite detector. In particular, within conventional physics, the SL EBL model predicts a photon survival probability $P_{\rm CP} \simeq 3 \times 10^{- 6}$ at ${\mathcal E} = 15 \, {\rm TeV}$, $P_{\rm CP} \simeq 1 \times 10^{- 8}$ at ${\mathcal E} = 18 \, {\rm TeV}$, $P_{\rm CP}\simeq 3 \times 10^{- 96}$ at ${\mathcal E} = 100 \, {\rm TeV}$ and $P_{\rm CP} \sim 0$ at ${\mathcal E} = 251 \, {\rm TeV}$. Therefore, due to the high EBL absorption of photons with ${\cal E} > 10 \, \rm TeV$, their observation requires a high TeV luminosity produced by GRB 221009A, which is hardly explainable within conventional emission models.  

In the Letter\cite{ALPinGRB}, the inclusion of the oscillation of photons into axion-like particles (ALPs) in the presence of external magnetic fields solves this problem, providing an explanation why GRB 221009A has been observed beyond $10 \, \rm TeV$. Specifically, when photons behave as ALPs, they are {\it not} EBL absorbed, thus increasing the photon effective mean free path, as deeply discussed in the literature (see e.g. the review\cite{universe}). Instead, the Lorentz invariance violation (LIV) is unable to justify the detection of GRB 221009A at LHAASO energies, as shown in the Letter\cite{ALPinGRB}.

\section{A model for GRB 221009A with axion-like particles}

Axion-like particles (ALPs) are a general prediction of many theories extending the standard model of particle physics including superstring and superbrane theories\cite{r2012}. ALPs are neutral very light spin-zero pseudo-scalar bosons with mass $m_a$, whose primary interaction with photons with coupling $g_{a \gamma \gamma}$ is described by the Lagrangian 
\begin{equation}
{\cal L}_{a \gamma \gamma} = - \, \frac{1}{4} \, g_{a \gamma \gamma} \, F_{\mu \nu} \, {\tilde{F}}^{\mu \nu} \, a = g_{a \gamma \gamma} \, {\bf E} \cdot {\bf B} \, a~,
\label{a2}
\end{equation}
with $a$ the ALP field and $F_{\mu \nu}$ (whose dual is ${\tilde{F}}^{\mu \nu}$) representing the electromagnetic tensor, where ${\bf E}$ and ${\bf B}$ denote its electric and magnetic components, respectively. In the present context, the effects of QED vacuum polarization\cite{hew1} and photon dispersion on the CMB\cite{raffelt2015} should also be taken into account.

From Eq.~(\ref{a2}), we observe that photons expressed by ${\bf E}$ can convert into ALPs in the presence of an external magnetic field ${\bf B}$ generating: (i) photon-ALP oscillations\cite{mpz,raffeltstodolsky} and (ii) the change of the polarization state of photons\cite{mpz,raffeltstodolsky}. Consequently, in astrophysical context, photon-ALP interaction causes observable effects both on spectra\cite{drm2007,simet2008,trgb2012,gtl,grdb} and on photon polarization\cite{bassan,galantiTheo,galantiPol,grtcClu,grtBlazar}.

The photons produced by GRB 221009A do not substantially convert into ALPs in the GRB jet, but can efficiently oscillate into ALPs in the galaxy hosting the GRB, as shown in the Letter\cite{ALPinGRB}. The host galaxy has been identified as a disc-like galaxy observed edge-on with the GRB placed in the nuclear region\cite{GRB221009Ahost}: we assume the two possibilities of (i) a typical spiral\cite{SpiralBrev} and (ii) a starburst similar to M82\cite{LopezRodriguez2021}. In both the previous situations, photon-ALP interaction turns out to be efficient. Photon-ALP conversion can also occur in the extragalactic space, where we use established models\cite{galantironcadelli20118prd,grjhea}. Specifically, we consider two possibilities concerning the extragalactic magnetic field strength $B_{\rm ext}$: (i) $B_{\rm ext} = 1 \, \rm nG$ (more likely\cite{kronbergleschhopp1999,furlanettoloeb2001}) with efficient photon-ALP conversion and (ii) the very conservative case $B_{\rm ext} < 10^{-15} \, \rm G$\cite{upbbext} with a negligible photon-ALP interaction. We observe no substantial modification of our results in the two cases concerning the value assumed by $B_{\rm ext}$. Finally, photon-ALP oscillation occurs in the Milky Way, where we employ established models concerning the behavior of the magnetic field\cite{jansonfarrar1} and of the electron number density\cite{yaomanchesterwang 2017}. After the evaluation of the photon-ALP beam transfer matrices in all the above-discussed crossed media, the final photon survival probability in the presence of photon-ALP oscillations $P_{\rm ALP}$ can be assessed\cite{ALPinGRB}.

\begin{figure}[h]
\centerline{\includegraphics[width=0.37\linewidth]{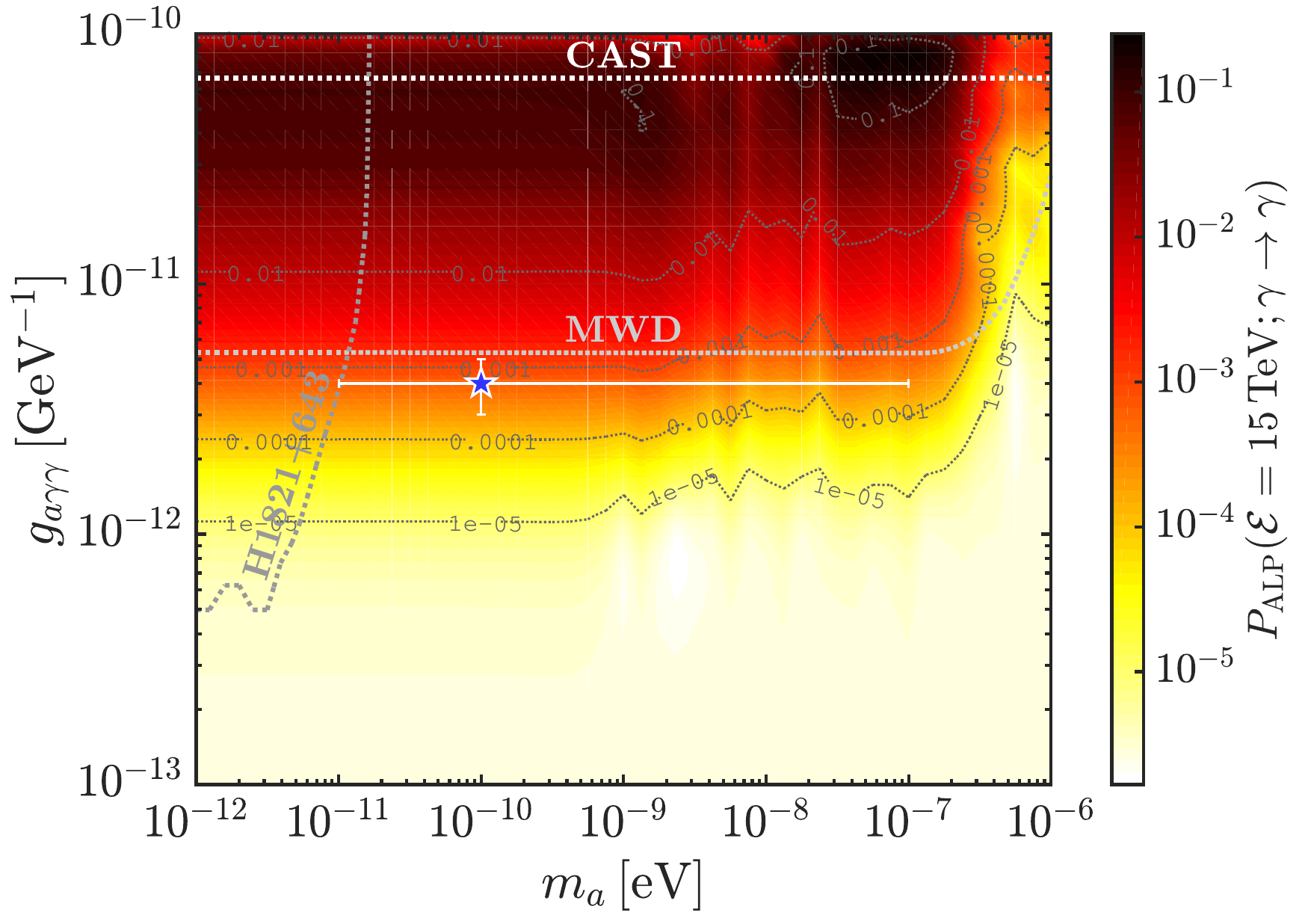}\includegraphics[width=0.33\linewidth]{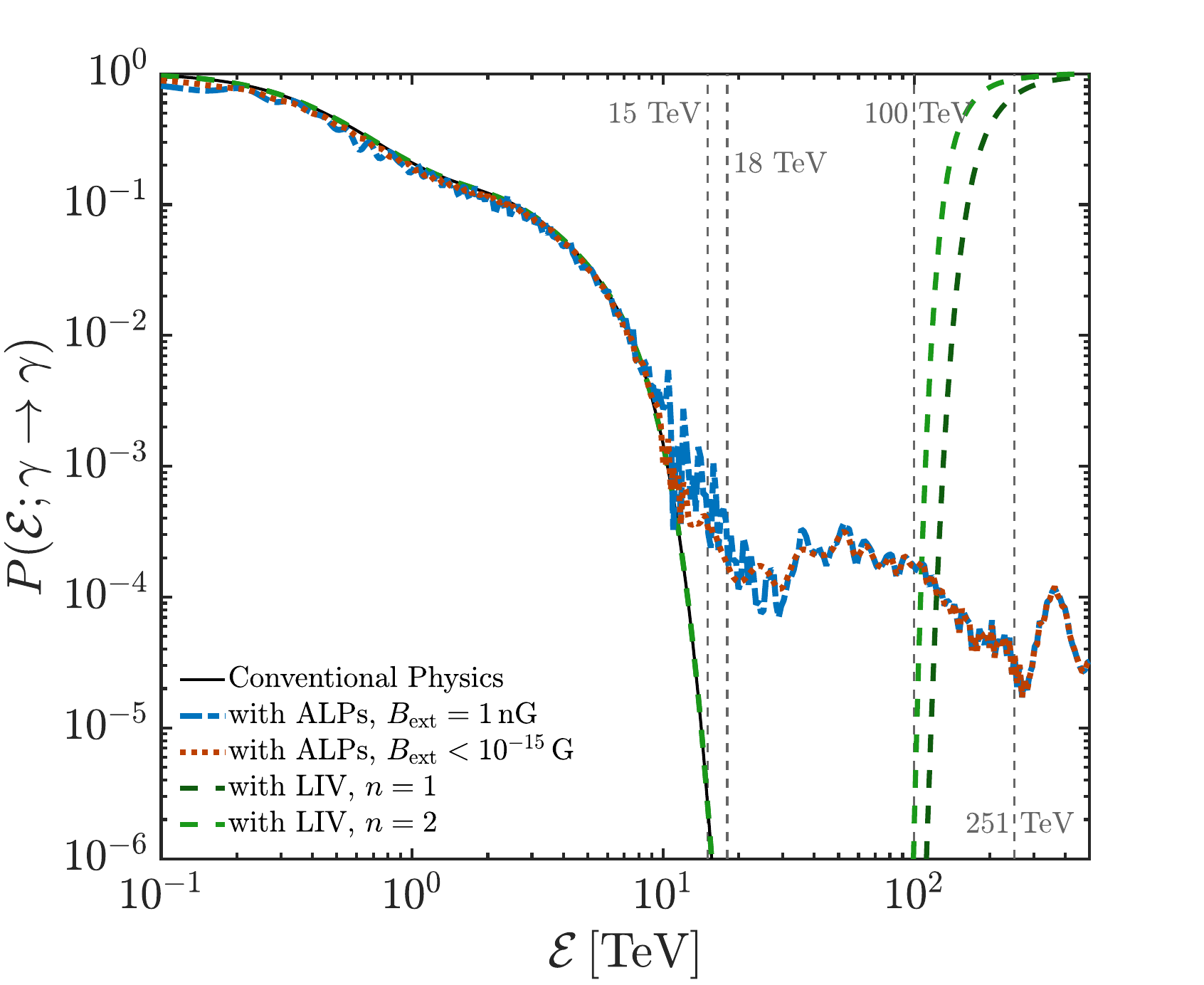}\includegraphics[width=0.33\linewidth]{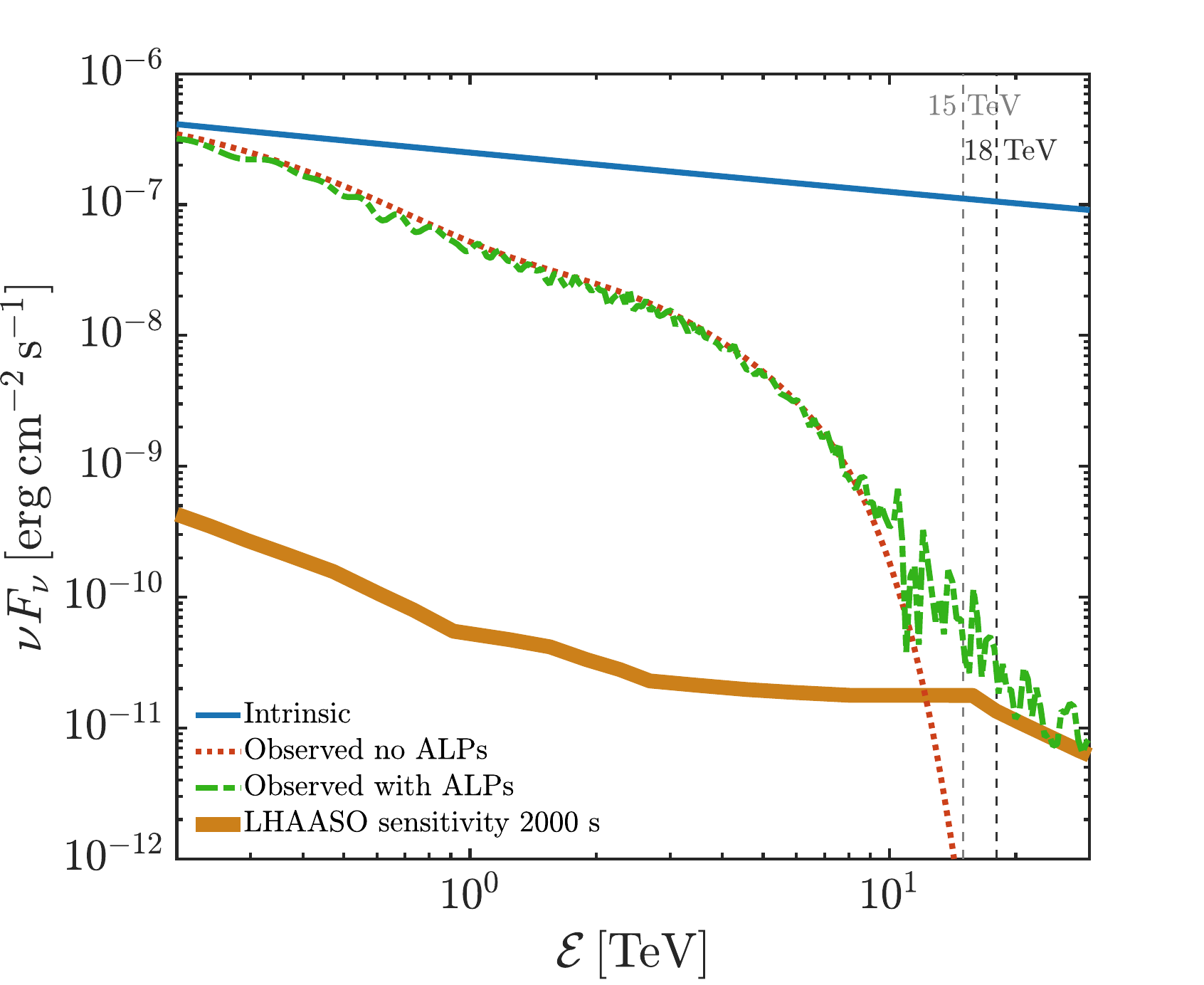}}
\caption[]{GRB 221009A: ALP parameter space and most reliable ALP bounds\cite{cast,limJulia,mwd} (left), photon survival probability in conventional physics and within photon-ALP interaction and LIV models (center), spectrum in conventional physics and within the photon-ALP interaction scenario along with LHAASO sensitivity (right).}
\label{figTOT}
\end{figure}

Our results are shown in Fig.~\ref{figTOT}, where we employ the SL EBL model\cite{saldanalopez} and we assume $B_{\rm ext}=1 \, \rm nG$ and a starburst galaxy hosting GRB 221009A.
In particular, in the left panel of Fig.~\ref{figTOT}, we show $P_{\rm ALP}$ at ${\cal E} = 15 \, \rm TeV$ (benchmark energy for the LHAASO event) as a function of $m_a$ in the interval $10^{- 12} \, {\rm eV} \lesssim m_a \lesssim 10^{- 6} \, {\rm eV}$ and of $g_{a\gamma\gamma}$ in the range $10^{- 13} \, {\rm GeV}^{- 1} \lesssim g_{a \gamma \gamma} \lesssim 10^{- 10} \, {\rm GeV}^{- 1}$ along with the most reliable ALP bounds to date [CAST experiment\cite{cast}, study of H1821+643\cite{limJulia} and of magnetic white dwarfs (MWD)\cite{mwd}]. The left panel of Fig.~\ref{figTOT} suggests us to take $m_a \simeq (10^{-11}-10^{-7}) \, \rm eV$ and $g_{a\gamma\gamma} \simeq (3-5) \times 10^{-12} \, \rm GeV^{-1}$ in order to maximize $P_{\rm ALP}$ around ${\cal E}=15 \rm \, TeV$ within all current most stringent ALP bounds. Correspondingly, along with $P_{\rm CP}$, the central panel of Fig.~\ref{figTOT} reports the behavior of $P_{\rm ALP}$ with respect to $\mathcal E$ assuming $m_a=10^{-10} \, \rm eV$ and $g_{a\gamma\gamma} = 4 \times 10^{-12} \, \rm GeV^{-1}$ as benchmark values according to two previous indications at the ALP existence\cite{trgb2012,grdb}, showing a substantial increase of $P_{\rm ALP}$ with respect to $P_{\rm CP}$, able to explain the LHAASO observation of GRB 221009A above $10 \, \rm TeV$. The GRB 221009A spectrum analyzed by LHAASO up to $7 \, \rm TeV$ once EBL-deabsorbed shows no cutoff\cite{grb221009aSpectrum}, which assures that the emitted spectrum can safely be extended up to $\sim 20 \, \rm TeV$ without any expected intrinsic deviation from a power law, as reported in the right panel of Fig.~\ref{figTOT}. The right panel of Fig.~\ref{figTOT} confirms that conventional physics can hardly explain the GRB 221009A spectrum above $\sim 10 \, \rm TeV$, while the photon-ALP interaction model developed in the Letter\cite{ALPinGRB} solves the problem (see also\cite{ALPinGRB2}), providing the strongest hint at ALP existence to date.

\section{Discussion and Conclusions}

Several models -- such as synchrotron self Compton (SSC) radiation or cascade emission from ultra-high energy protons\cite{gonzalez,das,zhao,sahu} --  have been proposed to explain the detection of GRB 221009A above $10 \, {\rm TeV}$\cite{LHAASO,grb221009aSpectrum,grb221009aSpectrum10TeV}, where photon absorption due to the EBL is extremely severe. However, such emission scenarios within conventional physics must assume {\it ad hoc} and contrived choices of the parameters without attaining a satisfactory description of GRB 221009A, when realistic EBL models, such as the SL one\cite{saldanalopez}, are considered.

The tension within conventional physics is solved by the photon-ALP oscillation scenario proposed in the Letter\cite{ALPinGRB}, thanks to which the EBL absorption is reduced to a level compatible with the LHAASO observation of GRB 221009A\cite{LHAASO,grb221009aSpectrum,grb221009aSpectrum10TeV}. The latter model works with ALP parameters within the most stringent bounds\cite{limJulia,mwd}.

The alternative scenario of Lorentz invariance violation (LIV) fails to justify the LHAASO detection of GRB 221009A, when current LIV bounds are taken into account, as reported in the central panel of Fig.~\ref{figTOT}, showing that LIV impact is effective for energies much larger than those pertaining to the LHAASO event (see the Letter\cite{ALPinGRB} and references therein for more details).

In conclusion, our findings imply a strong indication of ALP existence arising from GRB 221009A\cite{ALPinGRB} (see also\cite{ALPinGRB2}) with parameters aligned to those employed in two previous ALP hints\cite{trgb2012,grdb}, and which makes ALPs viable candidates for cold dark matter\cite{arias2012}. The recent data released by the LHAASO Collaboration concerning the GRB 221009A spectrum above $10 \, \rm TeV$\cite{grb221009aSpectrum10TeV} make our results about the solution of the problem thanks to photon-ALP interaction even more robust.

\section*{Acknowledgments}

The work of G. G. is supported by a contribution from Grant No. ASI-INAF 2023-17-HH.0 and by the INAF Mini Grant `High-energy astrophysics and axion-like particles', PI: Giorgio Galanti. The work of M. R. is supported by an INFN grant. The work by L. N. is partially supported by the INAF Mini Grant `Shock acceleration in Gamma Ray Bursts', PI: Lara Nava.

\end{document}